\documentclass[aps,preprint,showpacs,tightenlines]{revtex4}%
\usepackage{amsfonts}
\usepackage{amsmath}
\usepackage{amssymb}
\usepackage{graphicx}%
\begin{document}
\title{Critical currents in the BEC/BCS crossover regime}
\author{J. Tempere, J.T. Devreese}
\affiliation{TFVS, Departement Fysica, Universiteit Antwerpen, Universiteitsplein 1,
B-2610 Antwerpen, Belgium.}
\affiliation{\bigskip}
\keywords{BEC/BCS crossover, superfluidity, Fermi gas}
\pacs{03.75.-b, 03.75.Lm, 03.75.Ss}

\begin{abstract}
Both the trapping geometry and the interatomic interaction strength of a
dilute ultracold fermionic gas can be well controlled experimentally. When the
interactions are tuned to strong attraction, Cooper pairing of neutral atoms
takes place and a BCS superfluid is created. Alternatively, the presence of
Feshbach resonances in the interatomic scattering allows populating a
molecular (bound) state. These molecules are more tightly bound than the
Cooper pairs and can form a Bose-Einstein condensate (BEC). In this
contribution, we describe both the BCS and BEC regimes, and the crossover,
from a functional integral point of view. In this description, the properties
of the superfluid (such as vortices and Josephson tunneling) can be derived
and followed as the system is tuned from BCS the BEC. In particular, we
present results for the critical current of the superfluid through an optical
lattice and link these results to recent experiments with atomic bosons in
optical lattices.

\end{abstract}
\maketitle

\bigskip

\section{The ultracold dilute Fermi gas}

When a dilute Bose gas is cooled below the degeneracy temperature, the bosonic
atoms all condense in the same one-particle state and a Bose-Einstein
condensate forms. This has been convincingly demonstrated with magnetically
trapped, evaporatively cooled atomic gases for a multitude of atom species.
Moreover, magnetic or optical traps can be equally well loaded with fermionic
isotopes, such as $^{6}$Li or $^{40}$K. These atoms do not undergo
Bose-Einstein condensation, but fill up a Fermi sea, as has been demonstrated
through the observation of the Pauli blocking effect \cite{DeMarcoSCI285} and
through a measurement of the total energy of the Fermi gas \cite{DeMarcoPRL86}%
. Very soon after the observation of a degenerate Fermi sea of atoms,
researchers embarked upon the quest to achieve Cooper pairing in the dilute
Fermi gas. Indeed, for metals we know that the Fermi sea is unstable with
respect to Cooper pair formation. So, if the (neutral) atoms in the dilute gas
attract each other, a similar instability towards a paired state is to be expected.

\bigskip

The interatomic interactions in ultracold gases are remarkable for two
reasons. Firstly, the collisions between the atoms can be satisfactorily
characterized by a single number, the $s$-wave scattering length $a_{s}$. For
low-energy collisions, the effective interaction potential between atoms
becomes a contact potential, $V(\mathbf{r}-\mathbf{r}^{\prime})=g\delta
(\mathbf{r}-\mathbf{r}^{\prime}),$ where $g=4\pi\hbar a_{s}/m$ with $m$ the
mass of the atoms. The scattering length can be both positive (leading to
interatomic repulsion) or negative (attraction).

Secondly, this scattering length can be tuned by an external magnetic field
when a Feshbach resonance is present \cite{TiesingaPRA47}. This resonance
occurs when the energy of a bound (molecular) state in a closed scattering
channel becomes equal to the energy of the colliding atoms in the open
scattering channel. The different channels correspond here to different
hyperfine states of the trapped atoms, and the distance in energy between
these states can be tuned with a magnetic field.

\bigskip

In what follows, we will consider a trapped mixture of $^{40}$K atoms in the
$\left\vert 9/2,-7/2\right\rangle $ and $\left\vert 9/2,-9/2\right\rangle $
hyperfine states. This potassium isotope is fermionic, and the trapped states
display a Feshbach resonance at $B=202.1$ Gauss. When the scattering length is
tuned to a negative value, the atoms attract and Cooper pairs can form leading
to a \emph{BCS regime}. The critical temperature for Cooper pairing can be
raised by making the scattering length more strongly negative. When the
scattering length is large and positive, the molecular state in the closed
channel is populated, and molecules appear that can be Bose-Einstein condensed
(the \emph{BEC regime}). The adaptability of the scattering length allows
bringing the gas from the BCS regime into the BEC regime or vice versa, and
allows studying the interesting intermediate `crossover' regime.

\bigskip

The first experimental realization of superfluidity of a Fermi gas in the
molecular BEC\ regime came in 2003 \cite{ZwierleinPRL91}. A condensate of
molecules was convincingly observed. The detection of superfluidity in the BCS
regime however is much more subtle. In an initial experiment
\cite{OHaraSCI298}, the superfluid behavior was derived from the hydrodynamic
nature of the expansion of the cloud, as compared to a ballistic expansion
expected for a non-superfluid weakly-interacting Fermi gas \cite{MenottiPRL89}%
. However, this did not constitute unambiguous proof, since the Fermi gas was
in the strongly interacting regime. Subsequent experiments probed
superfluidity by mapping the pair density onto a molecular condensate density
\cite{RegalPRL92} or by spectroscopically measuring the gap
\cite{BartensteinPRL92}. Yet although these experimental methods clearly
demonstrate pairing, they do not unambiguously demonstrate superfluid behavior.

\bigskip%

\begin{figure}
[ptb]
\begin{center}
\includegraphics[
height=2.2026in,
width=3.1573in
]%
{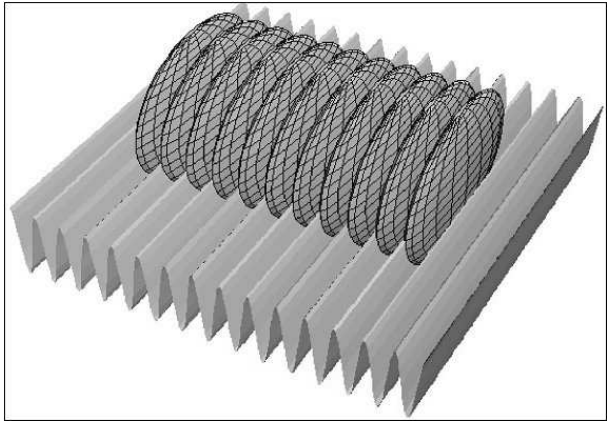}%
\caption{Two counterpropagating laser beams form a periodic potential for the
atoms. Such optical lattices can be loaded with quantum gases, forming a stack
of quasi-two dimensional clouds.}%
\label{optlatt}%
\end{center}
\end{figure}

The very recent observation of a lattice of quantized vortices in resonant
Fermi gases \cite{ZwierleinNAT435} constitutes the first clear demonstration
of superfluidity in the BEC/BCS regime. Observation of these vortices well in
the BCS regime may be difficult since the fermionic density penetrates in the
core of the vortex in the BCS regime, leading to a loss of contrast in direct
imaging \cite{BulgacJLTP138,TemperePRA71,ZwierleinNAT435}. Another possibility
to demonstrate superconductivity is though the observation of the Josephson
effect \cite{WoutersPRA70} in optical lattices. These optical lattices are
periodic potentials formed by two counterpropagating laser beams, for example
in the $z$-direction:%
\begin{equation}
V_{opt}(z)=sE_{R}\sin^{2}(2\pi z/\lambda), \label{Vopt}%
\end{equation}
where $\lambda$ is the laser wave length, $E_{R}=h^{2}/(2m\lambda^{2})$ is the
recoil energy, and $s$ is the laser intensity expressed in units of the recoil
energy. Typically, $s=1-20$, $\lambda=795$ nm. The atoms collect in the
valleys of the optical lattice and form a "stack of pancakes", illustrated in
Fig. \ref{optlatt}. Typically, there are on the order of a few 100 `pancakes'
with on the order of 1000 atoms each. When a superfluid is loaded in such an
optical lattice, the system corresponds to an array of Josephson junctions. In
such an array, the superfluid gas can propagate whereas the normal state gas
is pinned. This has already been demonstrated for bosonic atoms
\cite{CataliottiSCI293}, and has been predicted theoretically for fermionic
atoms \cite{WoutersPRA70,OrsoPRL93}.

In this contribution, we derive and discuss the critical Josephson current for
the flow of the superfluid component through an optical lattice. For this
purpose, we base ourselves on the path-integral description as applied by
Randeria and co-workers \cite{RanderiaPRB55,RanderiaPRL71} to the BEC/BCS
crossover model of high-T$_{c}$ superconductors. In section II, we give an
overview of the application of path-integrals to the system of ultracold
fermions, and in section III we present our results for the critical current.

\bigskip

\section{Path-integral treatment of the BEC/BCS crossover}

The partition function for the atomic Fermi gas is given by the functional
integral%
\begin{equation}
\mathcal{Z}=\int\mathcal{D}\bar{\psi}_{\mathbf{x},\tau,\sigma}\mathcal{D}%
\psi_{\mathbf{x},\tau,\sigma}\text{ }\exp\left\{  -\mathcal{S}/\hbar\right\}
\end{equation}
with an action%
\begin{align}
\mathcal{S}  &  =%
{\displaystyle\int\limits_{0}^{\hbar\beta}}
d\tau\int d\mathbf{x}\text{ }%
{\textstyle\sum_{\sigma}}
\bar{\psi}_{\mathbf{x},\tau,\sigma}\left(  \hbar\frac{\partial}{\partial\tau
}-\frac{\hbar^{2}}{2m}\nabla_{\mathbf{x}}^{2}-\mu\right)  \psi_{\mathbf{x}%
,\tau,\sigma}\nonumber\\
&  +%
{\displaystyle\int\limits_{0}^{\hbar\beta}}
d\tau\int d\mathbf{x}\text{ }g\bar{\psi}_{\mathbf{x},\tau,\uparrow}\bar{\psi
}_{\mathbf{x},\tau,\downarrow}\psi_{\mathbf{x},\tau,\downarrow}\psi
_{\mathbf{x},\tau,\uparrow}. \label{S1}%
\end{align}
The fermionic fields $\psi_{\mathbf{x},\tau},\bar{\psi}_{\mathbf{x},\tau}$ are
Grassman variables. The interaction potential, as discussed in the previous
section, is a contact potential with experimentally adjustable strength $g$.
The two hyperfine states are denoted by $\sigma=\uparrow,\downarrow$. The
functional integral over the Grassman variables can be performed analytically
only for an action that is quadratic in $\psi_{\mathbf{x},\tau},\bar{\psi
}_{\mathbf{x},\tau}$. In order to get rid of the quartic term in (\ref{S1}) we
perform a Hubbard-Stratonovic (HS) transformation, introducing auxiliary
bosonic fields $\bar{\Delta}_{\mathbf{x},\tau}$ and $\Delta_{\mathbf{x},\tau}$
:%
\begin{equation}
\mathcal{Z}\propto\int\mathcal{D}\bar{\psi}_{\mathbf{x},\tau,\sigma
}\mathcal{D}\psi_{\mathbf{x},\tau,\sigma}\int\mathcal{D}\bar{\Delta
}_{\mathbf{x},\tau}\mathcal{D}\Delta_{\mathbf{x},\tau}\text{ }\exp\left\{
-\mathcal{S}/\hbar\right\}
\end{equation}
with
\begin{align}
\mathcal{S}  &  =%
{\displaystyle\int\limits_{0}^{\hbar\beta}}
d\tau\int d\mathbf{x}\text{ }%
{\textstyle\sum_{\sigma}}
\bar{\psi}_{\mathbf{x},\tau,\sigma}\left(  \hbar\frac{\partial}{\partial\tau
}-\frac{\hbar^{2}}{2m}\nabla_{\mathbf{x}}^{2}-\mu\right)  \psi_{\mathbf{x}%
,\tau,\sigma}\nonumber\\
&  -%
{\displaystyle\int\limits_{0}^{\hbar\beta}}
d\tau\int d\mathbf{x}\text{ }\left(  \bar{\Delta}_{\mathbf{x},\tau}%
\psi_{\mathbf{x},\tau,\downarrow}\psi_{\mathbf{x},\tau,\uparrow}%
+\Delta_{\mathbf{x},\tau}\bar{\psi}_{\mathbf{x},\tau,\uparrow}\bar{\psi
}_{\mathbf{x},\tau,\downarrow}+\frac{\bar{\Delta}_{\mathbf{x},\tau}%
\Delta_{\mathbf{x},\tau}}{g}\right)  . \label{S2}%
\end{align}
Indeed, performing the functional integral over the HS fields $\bar{\Delta
}_{\mathbf{x},\tau}$,$\Delta_{\mathbf{x},\tau}$ in (\ref{S2}) brings us back
to (\ref{S1}). Our goal is an investigation of the superfluid properties of
the ultracold Fermi system. For a straightforward hydrodynamic interpretation
of the Hubbard-Stratonovic fields, it is advantageous to work with $\left\vert
\Delta_{\mathbf{x},\tau}\right\vert $ and $\theta_{\mathbf{x},\tau}$. These
are related to the original HS field by $\Delta_{\mathbf{x},\tau}=\left\vert
\Delta_{\mathbf{x},\tau}\right\vert \exp(i\theta_{\mathbf{x},\tau})$. We have
restricted the functional integral to $\bar{\Delta}_{\mathbf{x},\tau}%
=(\Delta_{\mathbf{x},\tau})^{\ast}$ without neglecting any field
configurations of importance to the final result. The hydrodynamic
interpretation of $\left\vert \Delta_{\mathbf{x},\tau}\right\vert ^{2}$ is the
density of fermion pairs, whereas $\hbar\mathbf{\nabla}_{\mathbf{x}}%
\theta_{\mathbf{x},\tau}/m=\mathbf{v}_{\mathbf{x},\tau}$ can be interpreted as
the superfluid velocity field. Performing this change of variables in the
functional integral yields%
\begin{equation}
\mathcal{Z}\propto\int\mathcal{D}\bar{\psi}_{\mathbf{x},\tau,\sigma
}\mathcal{D}\psi_{\mathbf{x},\tau,\sigma}\int\mathcal{D}\left\vert
\Delta_{\mathbf{x},\tau}\right\vert \mathcal{D}\theta_{\mathbf{x},\tau}\text{
}\exp\left\{  -\mathcal{S}/\hbar\right\}  , \label{Z3}%
\end{equation}
with%
\begin{align}
\mathcal{S}  &  =%
{\displaystyle\int\limits_{0}^{\hbar\beta}}
d\tau\int d\mathbf{x}\text{ }\bar{\psi}_{\mathbf{x},\tau,\sigma}\left(
\hbar\frac{\partial}{\partial\tau}-\frac{\hbar^{2}}{2m}\nabla_{\mathbf{x}}%
^{2}-\frac{1}{2}\mathbf{v_{\mathbf{x},\tau}}\cdot i\hbar\mathbf{\nabla
}_{\mathbf{x}}-\mu\right. \nonumber\\
&  \qquad\qquad\qquad+\left.  \frac{i\hbar}{2}\frac{\partial\theta
_{\mathbf{x},\tau}}{\partial\tau}-\frac{1}{4}\left(  i\hbar\mathbf{\nabla
}_{\mathbf{x}}\cdot\mathbf{v}_{\mathbf{x},\tau}\right)  +\frac{1}%
{8}m\mathbf{v}_{\mathbf{x},\tau}^{2}\right)  \psi_{\mathbf{x},\tau,\sigma
}\nonumber\\
&  -%
{\displaystyle\int\limits_{0}^{\hbar\beta}}
d\tau\int d\mathbf{x}\text{ }\left(  \left\vert \Delta_{\mathbf{x},\tau
}\right\vert \psi_{\mathbf{x},\tau,\downarrow}\psi_{\mathbf{x},\tau,\uparrow
}+\left\vert \Delta_{\mathbf{x},\tau}\right\vert \bar{\psi}_{\mathbf{x}%
,\tau,\uparrow}\bar{\psi}_{\mathbf{x},\tau,\downarrow}+\frac{\left\vert
\Delta_{\mathbf{x},\tau}\right\vert ^{2}}{g}\right)  . \label{S3}%
\end{align}
De Palo et al. \cite{DePaloPRB60} suggest at this point to introduce
additional collective quantum variables to extract the fermionic density.
However, care must be taken, since when additional collective quantum fields
are present the problem of double-counting poses itself \cite{KleinertFP26},
and variational perturbation theory has to be applied to avoid double-counting
\cite{KleinertAP266}. However, in the present case it is not necessary to
explicitly introduce the additional collective variables to obtain information
about the atomic density profile \cite{PelsterPC}. In (\ref{Z3}) the
integration over the fermionic variables can be taken, leading to%
\begin{equation}
\mathcal{Z}\propto\int\mathcal{D}\left\vert \Delta_{\mathbf{x},\tau
}\right\vert \mathcal{D}\theta_{\mathbf{x},\tau}\text{ }\exp\left\{
-\mathcal{S}_{\text{eff}}/\hbar\right\}
\end{equation}
with an effective action
\begin{equation}
\mathcal{S}_{\text{eff}}=-\hbar\operatorname*{tr}\left[  \ln\left(
\frac{-\mathbb{G}^{-1}}{\hbar}\right)  \right]  -%
{\displaystyle\int\limits_{0}^{\hbar\beta}}
d\tau\int d\mathbf{x}\text{ }\frac{\left\vert \Delta_{\mathbf{x},\tau
}\right\vert ^{2}}{g}. \label{S5}%
\end{equation}
where the inverse propagator can be written as the sum of an inverse `free
fermion propagator' and a term arising from the superfluidity:%
\begin{equation}
-\mathbb{G}^{-1}=-\mathbb{G}_{0}^{-1}+\mathbb{S}%
\end{equation}
The inverse free fermion propagator is%
\begin{equation}
-\mathbb{G}_{0}^{-1}=\sigma_{0}\left(  \hbar\frac{\partial}{\partial\tau
}\right)  +\sigma_{3}\left(  -\frac{\hbar^{2}}{2m}\mathbf{\nabla}_{\mathbf{x}%
}^{2}-\mu\right)  .
\end{equation}
and the superfluid part of the propagator can be written as%
\begin{align}
\mathbb{S}  &  =\sigma_{0}\left(  -\frac{1}{2}\mathbf{v_{\mathbf{x},\tau}%
}\cdot i\hbar\mathbf{\nabla}_{\mathbf{x}}\right)  -\sigma_{1}(\hbar\left\vert
\Delta_{\mathbf{x},\tau}\right\vert )\nonumber\\
&  +\sigma_{3}\left(  \frac{i\hbar}{2}\frac{\partial\theta_{\mathbf{x},\tau}%
}{\partial\tau}-\frac{1}{4}i\hbar\mathbf{\nabla}_{\mathbf{x}}\cdot
\mathbf{v}_{\mathbf{x},\tau}+\frac{1}{8}m\mathbf{v}_{\mathbf{x},\tau}%
^{2}\right)  .
\end{align}
In these expressions, $\sigma_{0}...\sigma_{3}$ are the Pauli matrices. Note
that if we have an external potential $V_{ext}(\mathbf{x)}$ present, for
example the optical potential or the magnetic trap, this appears in
$-\mathbb{G}_{0}^{-1}$ as an extra term $+\sigma_{3}V_{ext}(\mathbf{x)}%
$\textbf{.} The effective action (\ref{S5}) depends on the fields $\left\vert
\Delta_{\mathbf{x},\tau}\right\vert ,\theta_{\mathbf{x},\tau}$ and
$\rho_{\mathbf{x},\tau},\zeta_{\mathbf{x},\tau}$. For the former, a saddle
point approximation is usually made. For example, a good saddle point form
when no vortex is present is \cite{RanderiaPRL71,RanderiaPRB55}:%
\begin{equation}
\left\{
\begin{array}
[c]{c}%
\left\vert \Delta_{\mathbf{x},\tau}\right\vert =\Delta\\
\theta_{\mathbf{x},\tau}=\text{constant.}%
\end{array}
\right.  \tag{SP1}\label{SP1}%
\end{equation}
The value of the constant for the phase is irrelevant, and the value of
$\Delta$ can be extracted by extremizing the effective action $\delta
\mathcal{S}_{\text{eff}}/\delta\Delta=0$. This yields the well-known \emph{gap
equation }in the case of neutral atoms interacting through a contact
potential. Alternatively, we proposed in Ref. \cite{TemperePRA71} to use a
different saddle point approximation to investigate the case of a fermionic
superfluid containing a vortex parallel to the $z$-axis:%
\begin{equation}
\left\{
\begin{array}
[c]{c}%
\left\vert \Delta_{\mathbf{x},\tau}\right\vert =\Delta_{r}\\
\theta_{\mathbf{x},\tau}=\phi\text{.}%
\end{array}
\right.  \tag{SP2}\label{SP2}%
\end{equation}
Here, $\phi$ is the angle around the $z$-axis, and $r$ is the distance to the
$z$-axis. Again, a gap equation can be derived for $\Delta_{r}$ by extremizing
the action - this gap equation yields a gap that depends on the distance to
the vortex line (the $z$-axis). Fixing the total number of fermions yields a
number equation in which the local density of fermions can be identified straightforwardly.

\begin{figure}
[ptb]
\begin{center}
\includegraphics[
height=2.4193in,
width=3.0768in
]%
{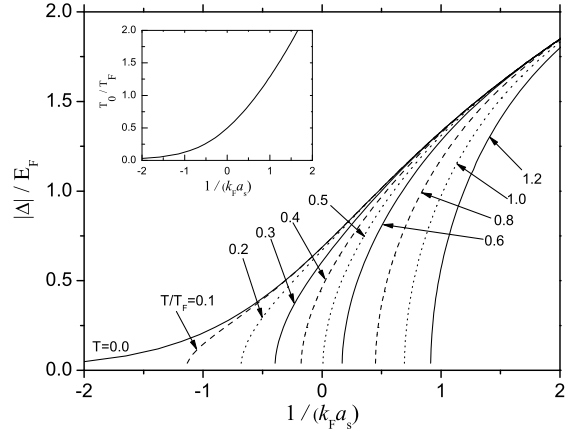}%
\caption{The saddle point value $\Delta~$is shown as a function of the
interaction strength parameter $1/(k_{F}a_{s})$ for different values of the
temperature. In the BCS regime this corresponds to the BCS gap and it vanishes
at the critical temperature. The inset shows the temperature at which
$\Delta=0$. In the BEC regime, fluctuations around the saddle point need to be
taken into account to obtain the correct critical temperature
\cite{RanderiaPRL71,PeraliPRL92,HuCM}.}%
\label{figdeltatemp}%
\end{center}
\end{figure}

Consider first the simplest saddle point approximation, (\ref{SP1}). The
saddle point result for the action in this case is
\begin{equation}
\mathcal{S}_{\text{sp1}}=\frac{\left\vert \Delta\right\vert ^{2}}{g}\left.
-2\int\dfrac{d\mathbf{k}}{(2\pi)^{3}}\ln\left[  2\cosh\left(  \dfrac{\beta}%
{2}\sqrt{\left(  \dfrac{k^{2}}{2m}-\mu\right)  ^{2}+\left\vert \Delta
\right\vert ^{2}}\right)  \right]  \right\}  .
\end{equation}
Two unknowns are the chemical potential $\mu$ and the value of constant
$\Delta$, the gap. The chemical potential is obtained by fixing the particle
density. In the BCS limit, $\mu\rightarrow E_{F}$ whereas in the BEC limit,
the chemical potential goes to the binding energy of the molecule,
$\mu\rightarrow\hbar^{2}/(ma_{s}^{2})$. In the intermediate regime, there is a
smooth crossover between the two limiting values. The gap $\Delta$ is found by
extremizing the saddle point action, $\delta\mathcal{S}_{\text{sp1}}%
/\delta\Delta=0$. The result is shown for different temperatures in figure
\ref{figdeltatemp}. At temperature zero, the gap depends exponentially on the
scattering length as we expect from the BCS theory. As the temperature is
raised, the gap decreases, reaching zero at a certain temperature. In the BCS
limit, the superfluidity is destroyed by breaking up Cooper pairs, so the
critical temperature corresponds to the temperature where $\Delta=0$. However,
in the BEC limit, superfluidity is destroyed through phase fluctuations, and
one cannot extract the critical temperature from the results shown in figure
\ref{figdeltatemp}. It becomes necessary to include fluctuations around the
saddle point value (\ref{SP1}) and expand the effective action up to second
order in these fluctuations around the saddle point value. This second-order
expansion yields an action that is quadratic in the fluctuation variables and
that can be integrated analytically. For fluctuations around the saddle point
(\ref{SP1}) this was done by Randeria and co-workers, who obtained a corrected
value of the critical temperature that in the BEC limit becomes independent of
$1/(k_{F}a_{s})$. More recently, the effects of fluctuations in the superfluid
regime, in the context of a diagrammatic expansion of the thermodynamic
potential in refs. \cite{PeraliPRL92,HuCM}.%

\bigskip

\bigskip

\section{Determination of the critical current}

\subsection{Effective action in the optical potential}

The path-integral method outlined in the previous section has been applied
before to describe vortices in a superfluid Fermi gas \cite{TemperePRA71} and
to describe the propagation of a superfluid Fermi gas in an optical potential
\cite{WoutersPRA70}. When an optical lattice (\ref{Vopt}) is present along the
$z$-direction, we can decouple the (free) motion in the $x,y$-plane from the
(tunneling) motion in the $z$-direction. To make this decoupling clear in the
notations, we write the partition function of the system as%
\begin{equation}
\mathcal{Z}=\int\mathcal{D}\bar{\psi}_{j,\sigma}(r)\mathcal{D}\psi_{j,\sigma
}(r)\text{ }\exp\left\{  -\mathcal{S}/\hbar\right\}  ,
\end{equation}
with $r=\{x,y,\tau\}$ and
\begin{equation}
\mathcal{S}=\sum_{j}\mathcal{S}_{j}+\sum_{j}\mathcal{S}_{j\rightarrow
j+1}^{tunnel}.
\end{equation}
The action functional for the gas in layer $j$ separately is given by
\begin{align}
\mathcal{S}_{j} &  =%
{\displaystyle\int\limits_{0}^{\hbar\beta}}
d\tau\int d\mathbf{x}\text{ }%
{\textstyle\sum_{\sigma}}
\bar{\psi}_{j,\sigma}(r)\left(  \hbar\frac{\partial}{\partial\tau}-\frac
{\hbar^{2}}{2m}\nabla_{\mathbf{x}}^{2}-V_{ext}(j)-\mu\right)  \psi_{j,\sigma
}(r)\nonumber\\
&  +%
{\displaystyle\int\limits_{0}^{\hbar\beta}}
d\tau\int d\mathbf{x}\text{ }g\bar{\psi}_{j,\uparrow}(r)\bar{\psi
}_{j,\downarrow}(r)\psi_{j,\downarrow}(r)\psi_{j,\uparrow}(r).
\end{align}
This is the two-dimensional version of the action functional (\ref{S1}),
supplemented with a layer index. Moreover, there is an external potential
$V_{ext}(j)$ acting on each layer. This can a parabolic potential in addition
to the optical potential itself. The tunneling of atoms from one layer to
another is described by%
\begin{equation}
\mathcal{S}_{j\rightarrow j+1}^{tunnel}=%
{\displaystyle\int\limits_{0}^{\hbar\beta}}
d\tau\int d\mathbf{x}\text{ }t_{1}\sum_{\sigma}\left[  \bar{\psi}_{j,\sigma
}(r)\psi_{j+1,\sigma}(r)+\bar{\psi}_{j+1,\sigma}(r)\psi_{j,\sigma}(r)\right]
,
\end{equation}
where the tunneling energy $t_{1}$ to bring an atom from one well of the
optical potential to the next was derived in Ref. \cite{MartikainenPRA68}:
\begin{equation}
t_{1}=sE_{R}\left[  \frac{\pi^{2}}{4}-1\right]  e^{-\sqrt{s}(\pi/2)^{2}}.
\end{equation}
For this particular decomposition of the action functional in intralayer
contributions and tunneling contributions, we can perform the same analysis as
described in the previous section. A Hubbard-Stratonovich transformation gets
rid of the four-operator term and introduces the HS fields $\left\vert
\Delta_{j}\right\vert ,\theta_{j}$, after which the integration over fermionic
variables is performed. The final result for the effective action can again be
written as the sum of contributions independent of $t_{1}$ and tunneling
contributions:%
\begin{equation}
\mathcal{S}_{\text{eff}}=\sum_{j}\mathcal{S}_{\text{eff,}j}+\sum
_{j}\mathcal{S}_{\text{eff, }j\rightarrow j+1}^{tunnel}.\label{Sopt}%
\end{equation}
The effective action for layer $j$ is
\begin{equation}
\mathcal{S}_{\text{eff,}j}=-\hbar\operatorname*{tr}\left[  \ln\left(
\frac{-\mathbb{G}_{\text{sp}}^{-1}}{\hbar}\right)  \right]  -%
{\displaystyle\int\limits_{0}^{\hbar\beta}}
d\tau\int d\mathbf{x}\text{ }\frac{\left\vert \Delta_{j}\right\vert ^{2}}%
{g}\label{Sopt-a}%
\end{equation}
with
\begin{equation}
-\mathbb{G}_{\text{sp}}^{-1}=\sigma_{0}\left(  \hbar\frac{\partial}%
{\partial\tau}\right)  +\sigma_{3}\left(  -\frac{\hbar^{2}}{2m}\mathbf{\nabla
}_{\mathbf{x}}^{2}-V_{ext}\left(  j\right)  -\mu\right)  -\sigma_{1}%
(\hbar\left\vert \Delta_{j}\right\vert ).
\end{equation}
The tunneling contributions in the effective action can be treated
perturbatively. In that framework, the saddle-point values $\left\vert
\Delta_{j}\right\vert $ can be extracted from the gap equation of each layer
separately, and the chemical potential $\mu$ is obtained from the number
equation. In each layer $j$, there is an `effective' chemical potential
$V_{ext}\left(  j\right)  +\mu$ fixing the local density $\rho_{j}$ in layer
$j$. Based on these results for the layers, the lowest-order perturbative
expansion of the action with respect to the tunneling part ($t_{1}$) yields%
\begin{equation}
\mathcal{S}_{\text{eff, }j\rightarrow j+1}^{tunnel}=-%
{\displaystyle\int\limits_{0}^{\hbar\beta}}
d\tau\int d\mathbf{x}\text{ }T_{j\rightarrow j+1}\cos[\theta_{j+1}-\theta
_{j}]\label{Sopt-b}%
\end{equation}
with,
\begin{equation}
T_{j\rightarrow j+1}=\frac{t_{1}^{2}\rho_{j}}{2\pi\hbar^{2}\rho_{j}/m+E_{b}},
\end{equation}
where $E_{b}$ is the binding energy of the molecule. This molecular binding
energy appears through the gap equations and can be derived from scattering
theory in reduced dimensionality. It is given by \cite{PetrovPRA64}:
\begin{equation}
E_{b}^{\text{2D}}=0.583\sqrt{s}E_{R}\exp\left(  \frac{1}{\sqrt{2\pi}}%
\frac{\lambda}{s^{-1/4}a_{s}}\right)  .
\end{equation}
It is important to note that the binding energy depends on the intensity and
wavelength of the lasers generating the optical potential. More intense laser
beams or smaller wavelengths confine the gas more strongly in the optical
lattice and alter the binding energy of the resonant molecules. A more
detailed determination of the molecular binding energy in an optical lattice,
taking into account molecules formed from atoms in neighboring lattice sites,
is given in Ref. \cite{WoutersCM}.

\subsection{Coupled density-phase equations}

The equations of motion for the remaining variables (density $\rho_{j}$ and
phase $\theta_{j}$ in layer $j$) can be derived from the effective action
(\ref{Sopt})-(\ref{Sopt-b}) through the extremum conditions $\delta
S_{\text{eff}}/\delta\theta_{j}=0$ and the number equation. This leads to the
equations reported by the present authors and M. Wouters in Ref.
\cite{WoutersPRA70}:%
\begin{align}
\hbar\partial_{t}\frac{\rho_{j}\left(  x\right)  }{2} &  =-\frac
{\mathbf{\nabla}\theta_{j}\cdot\mathbf{\nabla}\rho_{j}}{4m}\nonumber\\
&  +T_{j,j-1}\sin\left(  \theta_{j}-\theta_{j-1}\right)  \nonumber\\
&  -T_{j+1,j}\sin\left(  \theta_{j+1}-\theta_{j}\right)  ,\label{eqcont}%
\end{align}
and
\begin{align}
-\hbar\partial_{t}\frac{\theta_{j}}{2} &  =\frac{\left[  \nabla\theta
_{j}\right]  ^{2}}{8m}+V_{ext}\left(  j\right)  -\mu\nonumber\\
&  -\frac{\partial T_{j+1,j}}{\partial\rho_{j}}\cos\left(  \theta_{j+1}%
-\theta_{j}\right)  \nonumber\\
&  -\frac{\partial T_{j,j-1}}{\partial\rho_{j}}\cos\left(  \theta_{j}%
-\theta_{j-1}\right)  .\label{eEuler}%
\end{align}
In these equations, we have introduced the possibility of applying an external
potential $V_{ext}(j)$ varying over the layers. Here, we investigate the case
with a constant phase difference $\theta_{j+1}-\theta_{j}=\Delta\theta$ and a
smoothly varying density $\rho_{j+1}\approx\rho_{j}$. This situation
corresponds to a uniform flow of superfluid through the lattice. Then
equations (\ref{eqcont}),(\ref{eEuler}) simplify to%
\begin{align}
\hbar\partial_{t}\frac{\rho_{j}}{2} &  =\frac{t_{1}^{2}}{2\pi\hbar^{2}\rho
_{j}/m+E_{b}}\rho_{j}\sin(\Delta\theta)\label{cont2}\\
-\hbar\partial_{t}\frac{\Delta\theta}{2} &  =\nabla_{j}\left[  V_{ext}\left(
j\right)  \right]  -\frac{t_{1}^{2}E_{b}}{[2\pi\hbar^{2}\rho_{j}/m+E_{b}]^{2}%
}\cos(\Delta\theta)\label{eul2}%
\end{align}
In the BEC case, $E_{b}\gg\hbar^{2}\rho_{j}/m$ and we retrieve the equations
describing a conventional Josephson junction array. However, on the BCS side,
the tunneling coefficients start to depend on $\rho_{j}$, as $E_{b}$ and
$\hbar^{2}\rho_{j}/m$ become comparable.

\subsection{Critical Josephson current and critical velocity}

Equation (\ref{cont2}) states that the current density $J$ is proportional to
$\sin(\Delta\theta)$. This is similar to the first Josephson equation,
\[
J=J_{c}\sin(\Delta\theta).
\]
The second equation, (\ref{eul2}), fixes the time derivative of the phase
difference to the differences in potential energy of the gas in the different
wells. Near $\Delta\theta=0,\pi/2$ equation (\ref{eul2}) yields the second
Josephson equation $-\hbar d(\Delta\theta)/dt=\Delta U$ where $\Delta U$ is
the difference in energy between consecutive wells for the molecules. From the
first Josephson equation we can extract a critical current by setting
$\Delta\theta=\pi/2,$ as has been done for bosonic atoms in the optical
lattice in Ref. \cite{SmerziPRL89}. This yields a critical current density for
Josephson tunneling from layer to layer:%
\begin{equation}
J_{c}=\frac{2}{\hbar}\frac{t_{1}^{2}}{2\pi\hbar^{2}\rho_{j}/m+E_{b}}\rho_{j},
\end{equation}
The layers are separated by a distance $\lambda/2$. From $J_{c}$ we can then
extract the critical velocity for the fermionic atoms through the optical
lattice,
\begin{equation}
v_{c}=\frac{t_{1}^{2}}{2\pi\hbar^{2}\rho_{j}/m+E_{b}}\frac{\lambda}{\hbar}.
\label{vc}%
\end{equation}
%

\begin{figure}
[ptb]
\begin{center}
\includegraphics[
height=3.0876in,
width=4.0324in
]%
{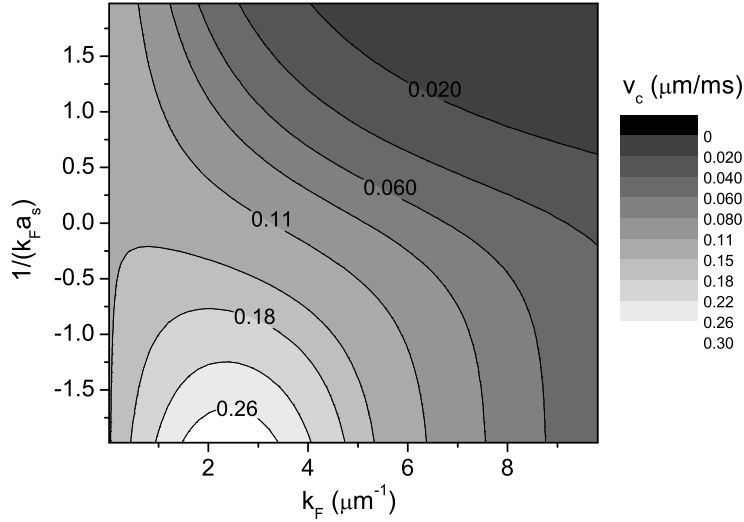}%
\caption{The critical Josephson velocity through the optical lattice, in
microns/ms, is shown as a function of the interaction strength parameter
$1/(k_{F}a_{s})$ and the Fermi wavevector of the 2D fermi gas in the valleys.
In the upper half ($a_{s}>0$) of the figure the BEC regime develops, and in
the lower half ($a_{s}<0$) the BCS regime appears. The optical lattice
parameters are $\lambda=795$ nm and $s=3$. The atom parameters correspond to
$^{40}$K atoms in the $\left\vert 9/2,-7/2\right\rangle $ and $\left\vert
9/2,-9/2\right\rangle $ hyperfine states, having a Feshbach resonance at
$B=202.1$ Gauss.}%
\label{figcritvel}%
\end{center}
\end{figure}

The critical velocity of the fermionic superfluid depends on the scattering
length $a_{s},$ via the binding energy of the Feshbach resonant molecule,
$E_{b}$. The critical velocity also depends on the density (or, equivalently,
the Fermi wave vector). In Figure \ref{figcritvel} we show the results for the
critical velocity (expressed in microns per millisecond), as a function of
$k_{F}$ and of the interaction parameter $1/(k_{F}a_{s})$. In the region
$1/(k_{F}a_{s})>0$ we are in the molecular BEC regime, and $E_{b}\gg\hbar
^{2}\rho_{j}/m$. The critical velocity in the BEC regime is roughly
proportional to $t_{1}^{2}/E_{b}$. In the region $1/(k_{F}a_{s})<0,$ the BCS
regime of Cooper pairs arises, and the result for the critical velocity
becomes nontrivial. For each fixed value of $1/(k_{F}a_{s})<0,$ there appears
a maximum as a function of $k_{F}$. This maximum occurs when $E_{b}%
\approx\hbar^{2}\rho_{j}/m$, minimizing the denominator in (\ref{vc}).

Although superfluid gases of bosonic atoms have already been studied in
optical lattices \cite{CataliottiSCI293,CataliottiNJP5}, superfluid Fermi
gases have to this moment not been loaded in optical lattices. Also no
molecular condensates have been placed in optical lattices. For atomic
condensates, a critical velocity could be determined experimentally
\cite{CataliottiNJP5}, and was found to vary between $0.2$ and $1.2$ $\mu$m/ms
for $^{87}$Rb atoms. This is comparable to the velocities that we predict for
(fermionic) $^{40}$K in the same $\lambda=795$ optical potential. Thus, the
superfluid regime of paired fermionic atoms in an optical lattice should be
accessible experimentally.

\section{Conclusions}

The path-integral description of ultracold fermionic atoms interacting through
a tunable contact potential allows to describe vortex configurations and other
non-ground state configurations through a judicious choice of saddle point. We
apply this formalism to the case of a fermionic gas in an optical potential.
When the fermionic gas is in the superfluid regime, the layers of gas in the
optical potential form a Josephson junction array. Equations of motion for the
density and phase in each layer are obtained and applied to the case where the
phase difference between consecutive layers is constant. This permits the
derivation of a critical velocity of the superfluid flow through the optical
potential. Although these results are strictly speaking derived for $T=0,$ in
the experiments the temperature can typically be brought down well below the
degeneracy temperature so that we believe our results will be relevant to the
experiments with optical lattices.

\begin{acknowledgments}
Stimulating discussions with H. Kleinert and A. Pelster are gratefully
acknowledged. J.T. acknowledges financial support from the FWO-Vlaanderen in
the form of a mandaat "Postdoctoraal Onderzoeker van het FWO-Vlaanderen". This
research has been supported financially by the FWO-V projects Nos. G.0435.03,
G.0306.00, the W.O.G. project WO.025.99N, the GOA BOF UA 2000 UA and the IUAP.
J.T. gratefully acknowledges support of the Special Research Fund of the
University of Antwerp, BOF\ NOI UA 2004.
\end{acknowledgments}

\end{document}